\documentclass[preprint,eqsecnum,aps,showpacs,nofootinbib]{revtex4}
\pdfoutput=1
\usepackage{bm,amssymb,amsmath,cancel,footnote,graphicx,subfigure}

\begin{document}

\count255=\time\divide\count255 by 60 \xdef\hourmin{\number\count255}
  \multiply\count255 by-60\advance\count255 by\time
 \xdef\hourmin{\hourmin:\ifnum\count255<10 0\fi\the\count255}

\newcommand{\xbf}[1]{\mbox{\boldmath $ #1 $}}

\title{Collider Signatures of the $N=3$ Lee-Wick Standard Model}

\author{Richard F. Lebed}
\email{Richard.Lebed@asu.edu}

\author{Russell H. TerBeek}
\email{r.terbeek@asu.edu}

\affiliation{Department of Physics, Arizona State University, 
Tempe, AZ 85287-1504}


\date{May 2012}

\begin{abstract}
  Inspired by the Lee-Wick higher-derivative approach to quantum field
  theory, Grinstein, O'Connell, and Wise have illustrated the utility
  of introducing into the Standard Model negative-norm states that
  cancel quadratic divergences in loop diagrams, thus posing a
  potential resolution of the hierarchy problem.  Subsequent work has
  shown that consistency with electroweak precision parameters
  requires many of the partner states to be too massive to be detected
  at the LHC\@.  We consider the phenomenology of a yet-higher
  derivative theory that exhibits three poles in its bare propagators
  (hence $N=3$), whose states alternate in norm.  We examine the
  interference effects of $W$ boson partners on LHC scattering cross
  sections, and find that the $N=3$ LWSM already makes verifiable
  predictions at 10~fb$^{-1}$ of integrated luminosity.
\end{abstract}


\maketitle
\thispagestyle{empty}

\newpage
\setcounter{page}{1}

\section{Introduction} \label{sec:intro}

Previous work by Grinstein, O'Connell, and
Wise~\cite{Grinstein:2007mp}, repurposing the seminal work of Lee and
Wick (LW)~\cite{Lee:1970iw}, has demonstrated the effectiveness of
introducing negative-norm states to cancel the quadratic divergences
endemic to the Standard Model (SM).  In this formulation, the LW
states amount to the additional solutions of a higher-derivative (HD)
field equation possessing two poles in its Feynman propagator.
Starting with
\begin{equation}
\mathcal{L}_{\rm HD}=-\frac{1}{2}\hat{\phi} \, \Box\hat{\phi}
-\frac{1}{2M^2}\hat{\phi} \, \Box^2\hat{\phi}
-\frac{1}{2}m^2\hat{\phi}^2 +\mathcal{L}_{\rm int}(\hat{\phi}) \, ,
\end{equation}
one obtains
\begin{equation}
\tilde{D}_{\rm HD}(p)=\frac{i}{p^2-m^2-p^4/M^2} \, ,
\end{equation}
which scales as $-p^{-4}$ at high energies, thereby improving the
convergence of Feynman diagrams in the HD formulation of the theory.
For $M^2\gg m^2$, the field $\hat{\phi}$ has poles at $p^2\approx m^2$
and $p^2\approx M^2$.  As in Ref.~\cite{Grinstein:2007mp}, one can
perform an auxiliary field redefinition to split the single field
$\hat{\phi}$ into two fields, one of positive and one of negative
norm:
\begin{equation}
\tilde{D}_{\rm SM}(p)\approx\frac{i}{p^2-m^2} \, , \ \
\tilde{D}_{\rm LW}(p)\approx\frac{-i}{p^2-M^2} \, ,
\end{equation}
which becomes exact in the limit $m^2/M^2\rightarrow 0$.  It was also
shown in~\cite{Grinstein:2007mp} that the heavy LW fields have the
interesting property of possessing negative decay widths, a feature
that figures prominently in the remainder of this work.

The primary theoretical motivation to study the LWSM is rooted in the
resolution of the hierarchy problem provided by higher-derivative
Lagrangians.  The level of attractiveness of this solution relies upon
whether these heavy LW states can be definitively observed at
colliders using present experimental capabilities ({\it i.e.}, the
LHC).  This line of research was initially conducted in
Ref.~\cite{Rizzo:2007ae}, in which the author analyzed, among other
processes, $pp\rightarrow W_i^+ + X \rightarrow \ell^+ + \nu_\ell +
X$, where $W_i^+$ refers to either a SM or LW virtual gauge boson, and
a LW mass of 1.5~TeV was assumed.  However, analyses of electroweak
parameters carried out in succession of
improvements~\cite{Alvarez:2008za,
Underwood:2008cr,Carone:2008bs,Chivukula:2010nw} (by scanning the LW
parameter space in~\cite{Alvarez:2008za}; by including only LW masses
for the fields most important for the hierarchy
problem~\cite{Carone:2008bs}; by using not just oblique parameters
$S$, $T$, but also the ``post-LEP'' parameters $W$,
$Y$~\cite{Underwood:2008cr}; by including bounds from the $Z b \bar b$
direct correction~\cite{Chivukula:2010nw}) reach the consensus
conclusion that the LW $W$ mass must be at least $\sim 3$~TeV in order
to maintain consistency with current electroweak precision tests
(EWPT), and even then only at the price of making LW fermion masses
substantially higher (as much as 10~TeV, according to the same
references).  More generic scenarios, in which all LW particles have
comparable masses, have typical bounds of $\agt 7$~TeV, which not only
places experimental signals out of reach for the LHC, but also
diminishes the value of LW states for providing a natural hierarchy
problem resolution.  Nevertheless, significant work on the
phenomenology of the LWSM at the LHC has been carried out in a number
of papers~\cite{Krauss:2007bz,Carone:2009nu,
Alvarez:2011ah,Figy:2011yu}.

However, the conventional LW approach studied to date provides only
the first nontrivial example in a series of higher-derivative theories
with an increasing number $N$ of propagator poles.  Naming the
original theory $N = 1$ and the conventional LW theory $N = 2$, one
may generate an $N = 3$ theory~\cite{Carone:2008iw}, or any higher
$N$,\footnote{The essential methods, at least at the
quantum-mechanical level, have been understood for a very long
time~\cite{Pais:1950za}.} each of which resolves the hierarchy
problem.  In particular, the Lagrangian for the scalar $N = 3$ theory
reads
\begin{equation}
\mathcal{L}_{N=3}=-\frac{1}{2}\hat{\phi} \, \Box\hat{\phi}
-\frac{1}{2M_2^2}\hat{\phi} \, \Box^2\hat{\phi}
-\frac{1}{2M_3^4}\hat{\phi} \, \Box^3\hat{\phi}
-\frac{1}{2}m^2\hat{\phi}^2 +\mathcal{L}_{\rm int}(\hat{\phi}) \, ,
\end{equation}
which gives rise to the propagator
\begin{equation}
\tilde{D}_{\rm HD}(p)=\frac{i}{p^2-m^2-p^4/M_2^2+p^6/M_3^4} \, .
\end{equation}
For $M_3 \gg M_2 \gg m$, this theory exhibits three well-separated
poles, in analogy to the $N = 2$ case discussed above.  However, when
one performs successive auxiliary field transformations to the
original HD Lagrangian, the new state so obtained is a heavy {\em
  positive}-norm field, with a positive decay width just like its SM
partner.  Work underway~\cite{LTprep} shows that this crucial
alternation of norm provides a substantial cancellation of the
complete LW contribution to the EWPT, and therefore significantly
relaxes the bounds on the LW masses; however, in anticipation of the
results of this detailed analysis, we consider here as a first step
whether the LW states themselves are easily discernible in realistic
LHC data, and whether they are easily distinguishable from other
beyond-SM (BSM) scenarios.  This work therefore closely follows the
approach of Ref.~\cite{Rizzo:2007ae}, which found the answers to these
questions for the LW $W$ to be {\em yes\/} and {\em no}, respectively;
we, on the other hand, show not only that the two LW $W$ states with
generic $\leq 2$~TeV masses are easily visible already with
10~fb$^{-1}$ of data, but also that the generic case produces a
spectrum distinguishable from that generated by other common BSM
scenarios.

The only assumptions essential to this analysis are the $+,-,+$
signature of the states and the $M_3>M_2>m$ hierarchy of masses.  The
construction of Ref.~\cite{Carone:2008iw} then shows that one can
build a canonically normalized Lagrangian of the form
\begin{equation}
\mathcal{L}=\displaystyle\sum_{i=1}^N(-1)^{i+1}
\left[ -\frac{1}{2}\phi^{(i)} \left( \Box + m_i^2 \right)
\phi^{(i)} \right] + \mathcal{L}_{\rm int} ( \{ \phi^{(i)} \} ) \, ,
\end{equation}
where $i=1$ refers to the SM state, $i=2$ refers to the negative-norm
LW state, $i=3$ refers to the heavy positive-norm LW state, and so on,
up to arbitrary $N$.

This paper is organized as follows: In Sec.~\ref{sec:methods}, we
outline the method of calculation for analyzing collider production of
$W_i$ states, and discuss the expected signals of such states from not
only LW, but Kaluza-Klein (KK) and other popular BSM scenarios as
well.  We present our results in Sec.~\ref{sec:results}, which
indicate definite LHC discovery potential, and offer conclusions in
Sec.~\ref{sec:conclusions}.  Technical details of the calculation are
relegated to the Appendix.

\section{Methods} \label{sec:methods}

We are primarily interested in the semileptonic process $pp\rightarrow
\ell^+ + \nu_\ell + X$, where the leptons are produced by an
intermediate $W_i^+$, and $X$ is an inclusive hadronic final state.
To leading order in weak interactions, one obtains the partonic-level
differential cross section~\cite{Rizzo:2007ae}
\begin{equation} \label{xsec}
  \frac{d^3\sigma}{d\tau\;dy\;dz}=K\frac{G_F^2M_W^4}{48\pi}
\displaystyle\sum_{qq'}|V_{qq'}|^2[SG^+_{qq'}(1+z^2)+2AG_{qq'}^-z],
\end{equation}
with the variables defined by
\begin{eqnarray} \label{S}
  S & \equiv & \displaystyle\sum_{ij}P_{ij}(C_iC_j)^\ell
(C_iC_j)^q(1+h_ih_j)^2 \, , \\
  A & \equiv & \displaystyle\sum_{ij}P_{ij}(C_iC_j)^\ell
(C_iC_j)^q(h_i+h_j)^2 \, , \\
  P_{ij} & \equiv & \hat{s}\frac{(\hat{s}-M_i^2)(\hat{s}-M_j^2)
    +\Gamma_i\Gamma_jM_iM_j}{[(\hat{s}-M_i^2)^2+\Gamma_i^2M_i^2]
    [i\rightarrow j]} \, . \label{S2}
\end{eqnarray}
Here, $K$ is a numerical factor $\simeq 1.3$ arising from NLO and NNLO
QCD corrections~\cite{Melnikov:2006kv}, and $z \equiv
\mathrm{cos}\;\theta^*$ is the center-of-momentum (CM) frame
scattering angle of parton $q$ into charged lepton $\ell^+$.  The
factors $P_{ij}$ represent cross terms between the allowed $W^+_i$
propagators (see the Appendix for a detailed description of the
calculation of $\Gamma$), and the $S$ (symmetric) and $A$
(antisymmetric) terms are the combinations of helicities $h_i$ and
couplings $C_i^{\ell,q}$ of the leptons and quarks that carry the
indicated parities with respect to the variable $z$.  The
$G_{qq'}^{\pm}$ are combinations of parton distribution functions
(PDFs)~\cite{Martin:2009iq}:
\begin{equation}
  G_{qq'}^{\pm}=q(x_a, \, M^2)\bar{q}'(x_b, \, M^2)\pm
q(x_b, \, M^2)\bar{q}'(x_a, \, M^2) \, ,
\end{equation}
where $q$ ($q^\prime$) are the PDFs associated with an up (down)-type
quark, the lepton invariant mass is $M^2 \equiv \hat{s}$, and
$x_{a,b}=\sqrt{\tau}e^{\pm y}$ are the parton longitudinal momentum
fractions, where $\tau \equiv \hat{s}/s$, and $y$ is the virtual gauge
boson rapidity.  At the level of the differential cross section, the
present work extends the calculation of Ref.~\cite{Rizzo:2007ae} by
introducing an additional heavy state with positive norm, and hence
with $\Gamma_{3}>0$.  One converts the differential cross section
Eq.~(\ref{xsec}) into a distribution in the transverse mass $M_T$,
obtained from $z = (1-M_T^2/M^2)^{1/2}$:
\begin{equation}
  \frac{d\sigma}{dM_T}=\displaystyle\int_{M_T^2/s}^1d\tau
  \int_{-Y}^{Y}dy\;J(z\rightarrow M_T)\frac{d^3\sigma}{d\tau\;dy\;dz}
  \, .
\end{equation}
The Jacobian factor $J(z\rightarrow M_T)=|dz/dM_T|
=\frac{M_T}{M^2}|1-M_T^2/M^2|^{-1/2}$ produces the peak structures
seen in the plots in the next section.  The new positive-norm state
produces an observable signal in the region near $M_T\approx M_3$,
with a distinctive sharp interference edge generated by the
off-diagonal terms in $P_{ij}$.

How do the predictions of the $N=3$ LWSM compare with those of other
theories featuring a heavy $W'$ boson?  The $W'$ could, in principle,
have arbitrary helicities and couplings to SM fields.  The
alternating-norm LW states offer just one such example amidst a
plethora of possibilities.  These include the Sequential Standard
Model (the SM with extra gauge bosons carrying the same couplings),
left-right symmetric models (such as Pati-Salam), and KK excitations
of the $W$ on a compactified $S^1/{Z}_2$ dimension
(see~\cite{Davoudiasl:2007cy} for a thorough list of references).  It
was shown in~\cite{Rizzo:2007ae} that the first two scenarios are
already clearly distinguishable from the $N=2$ LWSM, so we do not
consider them further.  The hypothetical KK excitations of the $W$, on
the other hand, demand a more careful treatment.

The most straightforward universal extra dimension
models~\cite{Barbieri:2004qk} already require $R^{-1}$ to exceed
several TeV, but alternative mechanisms allow $R^{-1}$ to be brought
down to scales that can be probed at the LHC\@.\footnote{KK
  excitations with masses beyond several TeV are directly observable
  at the LHC only with a much greater integrated luminosity than is
  presently available.}  Suppose, for example, that gauge and Higgs
bosons propagate in the bulk of the $S^1/Z_2$ dimension $y\in[0,\pi
R]$, but that leptons are localized
at $y=0$ and quarks are localized at $y=\pi
R$~\cite{ArkaniHamed:1999za}.  Such a mechanism can have a lower
compactification scale; as in Ref.~\cite{Rizzo:2007ae}, for sake of
argument we take it to be $R^{-1}\sim1.5$~TeV\@.  The $n^{\rm th}$
mode of the $W$ in the compactified dimension has a 5D wave function
of the form $\cos(ny/R)$.  Reading off the couplings from Eq.~(3) of
Ref.~\cite{Rizzo:1999br}, one sees that the localization of the quarks
at $y=\pi R$ forces their couplings in the 4D effective theory to be
$C_n^q=(-1)^{n}$ for the $n^{\rm th}$ KK excitation of the $W$.  Upon
making this change to Eq.~(\ref{S}), one finds that the KK excitations
can mimic the wrong-sign propagator of the LW bosons quite faithfully.
The only algebraic difference comes from the negative sign of
$\Gamma_{2}$, but due to the very narrow widths under consideration
(as calculated in the Appendix), the two models are potentially
virtually indistinguishable.  This ambiguity could not be resolved
within the framework of the $N=2$ LWSM, but we find that the $N=3$
LWSM yields starkly different predictions.

To illustrate this point, consider the mass term of the KK
excitations, with $\varphi_b$ being the bulk Higgs field with VEV
$|\varphi_b|$~\cite{Rizzo:1999br} and $g$ being the 5D gauge coupling:
\begin{equation} \label{KKmass}
  \mathcal{L}_{\rm mass}=\frac{1}{2}
  \left(\frac{n^2}{R^2}+2g^2|\varphi_b|^2\right)
  V_{\mu}^{(n)}V^{(n)\mu} \, .
\end{equation}
%
One sees that the squared masses of the KK excitations obey a specific
$n^2$ hierarchy; therefore, if the first excitation has a mass equal
to that of the $N=2$~$W$ boson, the mass of the second KK excitation is
pre-determined by the theory, whereas the masses of LW bosons can in
principle assume any positive values.  The spacing restriction is
especially apparent in the limit $R^{-1} \gg g |\varphi_b|$, in which
case the KK excitations have equally spaced masses
\begin{equation}
  m_{\rm KK} \approx \frac{n}{R} \, .
\end{equation}
In order for the KK and $N=3$ LWSM scenarios to be confused, either
the experimental sensitivities must be such that only one excitation
can be discerned in each case (which reduces to the situation
described in Ref.~\cite{Rizzo:2007ae}), or the spacing of the LW
partners matches by unfortunate chance the spectrum given by
Eq.~(\ref{KKmass}).  In this case, the natural next step would be to
look for the KK and LW partners of other particles and examine their
mass spectra for further evidence to distinguish the two
possibilities.  In general, however, we conclude that the generic
$N=3$ LWSM makes unique predictions for mass and coupling spectra that
cannot be mimicked by well-known alternative $W'$ theories.

\begin{figure}
  \centering
  \includegraphics[width=120mm]{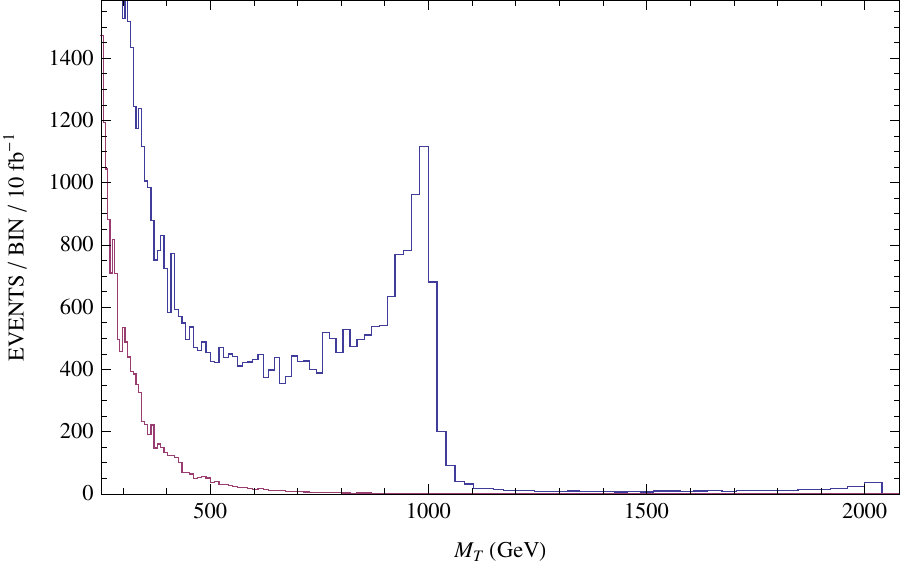}
  \includegraphics[width=120mm]{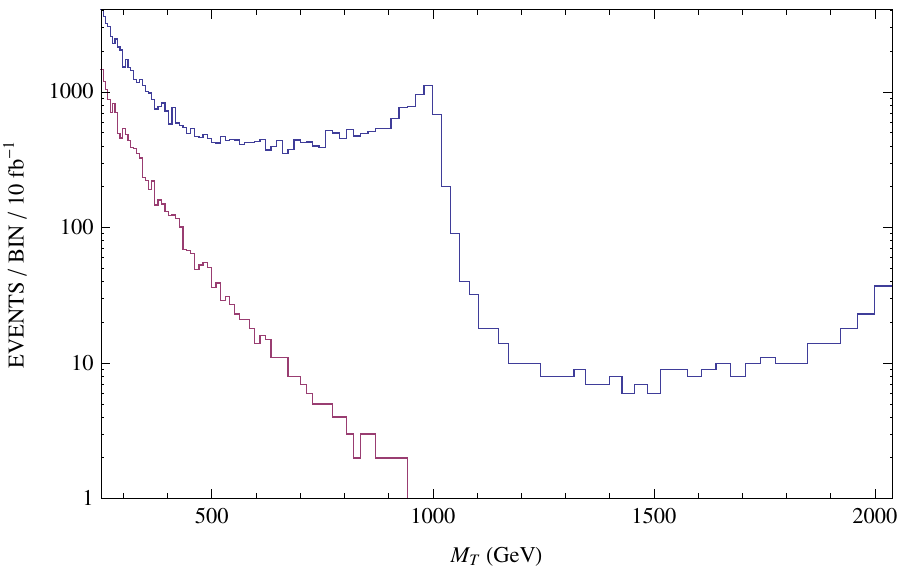}
  \caption{{\footnotesize Transverse mass distribution for the $N=3$
      LWSM (blue) and expected SM background (red), assuming masses
      $M_2=1$~TeV, $M_3=2$~TeV.  Both plots display the same data,
      though the lower plot has a log scale in order to emphasize the
      second Jacobian peak near the region $M_T\approx M_3$.  We
      employ a cut $|\eta_\ell|<2.5$ on the rapidity of the outgoing
      leptons, and smear the distribution by $\delta
      M_T/M_T\approx2\%$ to simulate the resolution of the ATLAS
      detector.}} \label{M2eq1M3eq2}
\end{figure}
\begin{figure}
  \centering
  \includegraphics[width=120mm]{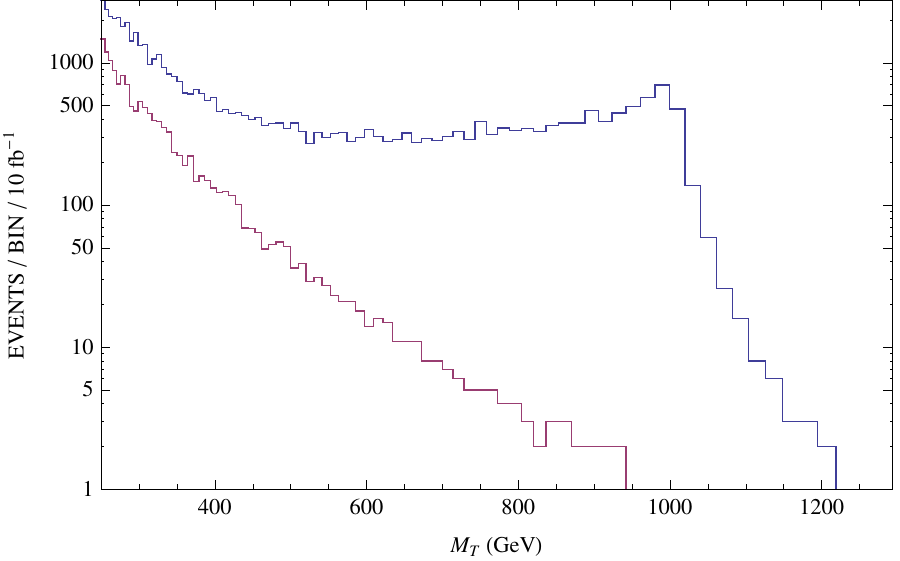}
  \caption{{\footnotesize Transverse mass distribution for $M_2 =
      1$~TeV, but now $M_3 = 5$~TeV.  Even with an integrated
      luminosity of 10~fb$^{-1}$, the second Jacobian peak is too weak
      to be discernible; the plots are truncated where the simulated
      number of events per bin drops below 0.5.}} \label{M2eq1M3eq5}
\end{figure}
\begin{figure}
  \centering
  \includegraphics[width=120mm]{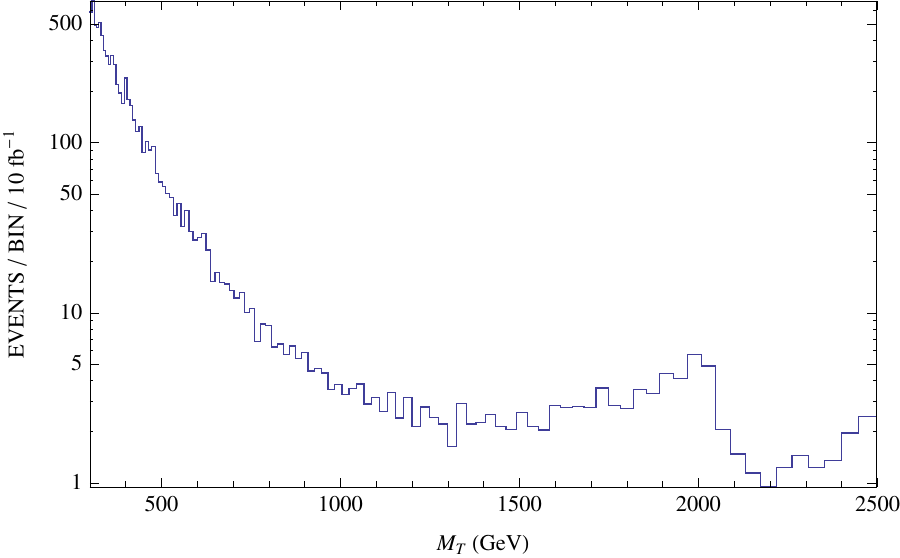}
  \caption{{\footnotesize Transverse mass distribution for $M_2 =
      2$~TeV, $M_3 = 2.5$~TeV, now at $\sqrt{s} = 8$~TeV and an
      integrated luminosity of 15~fb$^{-1}$.}}
  \label{15fb-1}
\end{figure}

\section{Results} \label{sec:results}

We start with the LHC inputs $\sqrt{s} = 7$~TeV and 10~fb$^{-1}$ of
integrated luminosity.  Assuming the masses $m \approx m_{W,{\rm
    SM}}=80.4$~GeV, $M_2 = 1$~TeV, $M_3 = 2$~TeV, a set chosen purely
for illustration, one then has all the necessary inputs to compute
transverse mass distributions in the LWSM\@.  We plot our results in
Fig.~\ref{M2eq1M3eq2}.  The most exciting feature is the statistically
robust Jacobian peak near $M_T\approx M_3$ at 10~fb$^{-1}$, which is
not only revealed as dozens of events that would not appear in the SM
or even the $N=2$ LWSM, but one that exhibits a very distinctive
morphology.  This result indicates that, for a sufficiently light
$W_3^{\pm}$ boson, the $N=3$ LWSM makes unambiguous predictions ({\it
  i.e.}, are highly unlikely to be confused with KK modes or even the
$N=2$ LW theory) that can be tested at the LHC, given a reasonable
10~fb$^{-1}$ of integrated luminosity.  However, we hasten to add that
nothing is special about the value $M_3=2$~TeV; in principle, the
$W_3^{\pm}$ could be significantly heavier, still resolving the
hierarchy problem (although likely creating tension with
EWPT~\cite{LTprep}), but evading detection any time within the next
decade.  Behavior that satisfies a theory well can provide a dull
phenomenology: A sufficiently heavy $W_3^{\pm}$ combined with a
lighter $W_2^{\pm}$ might still satisfy the EWPT constraints while
slipping past the range of detection.  Even for $M_3=5$~TeV this
effect is quite stark; see Fig.~\ref{M2eq1M3eq5} for a plot of the
transverse mass distribution in such a theory.  One can check that the
absence of a strong $M_3=5$~TeV signal is not due to the limited
integrated luminosity, but rather the limited CM energy.

One can also show that a 2012 LHC data set of 15~fb$^{-1}$ at
$\sqrt{s} = 8$~TeV allows the discovery bounds to be pushed even
further.  In Fig.~\ref{15fb-1} we present the expectation in such a
scenario for $M_2 = 2.0$~TeV and $M_3 = 2.5$~TeV, which provides
further evidence that a wide expanse of LW parameter space will be
probed in the near future at the LHC\@.  The crucial point of the
present work is that, regardless of the masses of the $W_i$ under
consideration, the alternating-sign metric of the LWSM plus the lack
of a specific mass spectrum functional dependence produces distinct
Jacobian peak structures that can be easily distinguished from those
of other theories.

\section{Conclusions} \label{sec:conclusions}

Our work in the $N=3$ LWSM shows that the presence of a heavy,
positive-norm state has observable consequences for physics at the
LHC\@.  It makes predictions above and beyond that of the conventional
$N=2$ LWSM, and moreover is clearly distinguishable from other heavy
$W'$ theories, most specifically the Kaluza-Klein excitations
associated with particles propagating along a compactified $S^1/Z_2$
dimension.  The novel contribution of our work is the second Jacobian
peak associated with the $N=3$ $W$ boson.  However, since little
information exists to constrain the value of $M_3$~\cite{LTprep}, our
choice $M_3=2$~TeV should be interpreted as one possibility of many.
We find that, for a sufficiently heavy choice of $M_3$ ($\sim 5$~TeV),
the expected signal for the $N=3$ theory does not rise significantly
above the SM background for 10~fb$^{-1}$ of integrated luminosity.

Finally, one might question the naturalness of the choice $N=3$, and
argue either that one has no good reason to stop at 3, or that
including as many as 2 LW partners is already excessive.  For
argument's sake, why would one settle with a theory of $N$ sets of
states when $N+1$ sets might do an even better job of solving the
hierarchy problem?  It is even possible (albeit unpalatable) for a
countably infinite tower of LW states to exist, alternating in the
signs of their norms, while still conspiring to cancel out divergences
in loop diagrams; the study of field equations of motion for
infinite-order propagator poles was addressed in
Ref.~\cite{Pais:1950za}, wherein the authors found that basic elements
such as the propagator could still be reliably constructed in such a
theory.  These increasingly heavy states would serve an important
theoretical role in the structure of the SM, but would lie far beyond
the reach of any collider in the foreseeable future, not to mention
creating a proliferation of new degrees of freedom.  Moreover, as
discussed in the Introduction, the $N = 2$ LWSM with lighter partner
masses is known to create significant tension with electroweak
precision constraints (particularly for custodial isospin violation),
but we anticipate a large degree of cancellation at the $N = 3$ level
between the opposite-norm LW partners.  For these reasons, we have
restricted our attention solely to the $N=3$ LWSM; the full range of
phenomenologically allowed possibilities awaits an analysis of the
precision constraints~\cite{LTprep}.

\appendix

\section{Diagonalization of the Gauge Boson and Quark Sectors and
  Calculation of Decay Widths}

\subsection{Gauge Boson Mass Diagonalization}

First address the question of diagonalizing gauge boson masses.  In a
LW theory with spontaneous symmetry breaking, one generically
encounters mass mixing terms between the SM and LW gauge bosons; one
must diagonalize this sector in order to construct gauge boson mass
eigenstates.  Start with the Higgs kinetic energy Lagrangian
\begin{equation}
\mathcal{L}_{\rm Higgs, kin}
=\eta_{ij}(\hat{D}_{\mu}H_i)^{\dagger}(\hat{D}^{\mu}H_j) \, ,
\end{equation}
where the metric $\eta_{ij}=\rm diag\{1,\;-1,\;1\}$ encodes the
opposite signs of the LW states, and the $H_i$ are given by
\begin{equation}
H_{1} = \left(\begin{array}{c} 0 \\ \frac{1}{\sqrt{2}}(v+h_1)
\end{array} \right) , \,\,\,\,\,
H_{2} = \left(\begin{array}{c} h_2^+ \\ \frac{1}{\sqrt{2}}
(h_2 + i P_2) \end{array} \right) ,
\,\,\,\,\,
H_{3} = \left(\begin{array}{c} h_3^+ \\ \frac{1}{\sqrt{2}}
(h_3 + i P_3) \end{array} \right) \, .
\end{equation}
Note that only $H_{1}$ carries a nonzero VEV\@.  The hat on $\hat D$
indicates an action on superfields containing both the SM field and
its LW partners.  For fields such as $H_i$ transforming under the
fundamental representation of the group with gauge bosons denoted by
$A^\mu$, one has $\hat{D}^{\mu}=\partial^{\mu}-ig\hat{A}^{\mu}$.  Upon
expanding the HD gauge superfield $\hat{A}^{\mu}$ in terms of its LW
components and resolving the resulting fields into mass eigenstates,
one finds
\begin{equation} \label{gaugesuper}
  \hat{A}^{\mu}=A_1^{\mu}-\frac{M_3}{\sqrt{M_3^2-M_2^2}}A_2^{\mu}
  +\frac{M_2}{\sqrt{M_3^2-M_2^2}}A_3^{\mu}\equiv
  \hat{\theta}_p A_p^{\mu}=\hat{\theta}_p V_{pq} A_{q,0}^{\mu} \, ,
\end{equation}
where the first equality is due to Eq.~(3.9) of
Ref.~\cite{Carone:2008iw}.  The correlation of SM and LW
fields~\cite{Carone:2008iw} with $m \equiv M_1 = 0$ is incorporated by
defining the vector
\begin{equation} \label{thetadef}
  \hat{\theta} \equiv \left\{ 1,\;-\frac{M_3}{\sqrt{M_3^2-M_2^2}},\;
  \frac{M_2}{\sqrt{M_3^2-M_2^2}} \right\} \, ,
\end{equation}
the mass eigenstates are denoted by a 0 subscript, and $V$ is a mixing
matrix whose origin is explained shortly.  Expanding the superfield
covariant derivative for the gauge fields using
Eq.~(\ref{gaugesuper}), one notes that in the absence of
spontaneous symmetry breaking, the fields $A_p^{\mu, a}$ are the
eigenstates with masses given by $M_p$.  However, the nonzero Higgs
VEV connects all terms quadratic in the SU(2) fields $W_p^{\mu, a}$
equally, mandating somewhat more effort to obtain the complete mass
eigenstates.  Keeping only the terms quadratic in $W^\pm$,
$\mathcal{L}_{\rm Higgs, kin}$ evaluated at its VEV gives the
additional mass contribution
%
\begin{equation}
\Delta {\cal L}_{\rm mass} = \left( \frac{g_2 v}{2} \right)^2
\hat{\theta}_r \hat{\theta}_s W_{r\mu}^+W_{s}^{-\mu} \, .
\end{equation}
One sees that the Higgs VEV introduces a contribution to the gauge
boson masses beyond the explicit masses $M_2$, $M_3$ generated by the
pure Yang-Mills Lagrangian of the theory.  Since the diagonalization
of the gauge sector entails solving the eigenvalue problem of a
$3\times3$ matrix, we introduce numerical matrices $V^+$ satisfying
$W_p^{+}=V^{+}_{pq} W_{0,q}^{+}$, where $W_{0,q}^{+}$ are the mass
eigenstates associated with the eigenvalues in the propagator factors
of Eq.~(\ref{S2}).

\subsection{Quark Mass Diagonalization}

For the quarks, begin with definitions of the purely left-handed
supermultiplets $T^T_L \equiv \left(t^{\vphantom{'}}_{L,1},
  t^{\vphantom{'}}_{L,2},t_{L,2}',
  t^{\vphantom{'}}_{L,3},t_{L,3}'\right)$ and $B^T_L \equiv
\left(b^{\vphantom{'}}_{L,1}, b^{\vphantom{'}}_{L,2},b_{L,2}',
  b^{\vphantom{'}}_{L,3},b_{L,3}'\right)$, and define $T_R^T$ and
$B_R^T$ analogously.  (Of course, $T$ and $B$ can refer to the
collection of all up- and down-type quarks, respectively.)  Here, the
unprimed LW fields possess the same quantum numbers as their SM
counterparts [{\it i.e.}, all $t_{L,i}$ transform as
$(\textbf{2},+\frac{1}{6})$ under SU(2)$\times$U(1)], whereas the
primed fields possess quantum numbers identical to the {\em
  unprimed\/} fields of {\em opposite\/} chirality, so that all
$t_{L,i}'$ (like $t^{\vphantom{'}}_{R,i}$) transform as
$(\textbf{1},+\frac{2}{3})$ and all $t_{R,i}'$ (like
$t^{\vphantom{'}}_{L,i}$) transform as $(\textbf{2},+\frac{1}{6})$.
This relationship can be clarified by examining the derivation of the
HDLW Lagrangian from the original $N=3$ HD
theory~\cite{Carone:2008iw}, but arises quite generally for BSM
theories in which fermions are permitted to possess explicit Dirac
mass terms.

Now examine the mass contribution to the Lagrangian, which allows all
SM and LW states to mix.  These terms are of the form
\begin{equation} \label{supermass}
  \mathcal{L}_{\rm mass}=-\overline{T}^{\vphantom{dagger}}_L \, \rho
  \mathcal{M}_t^{\dagger} T^{\vphantom{dagger}}_R
  -\overline{B}^{\vphantom{dagger}}_L \, \rho\mathcal{M}_b^{\dagger}
  B^{\vphantom{dagger}}_R+ {\rm h.c.} \, ,
\end{equation}
where the metric $\rho\equiv\mathrm{diag}\{1,-1,-1,1,1\}$ conveniently
encodes the opposite signs of the positive- and negative-norm states.
Not counting the Hermitian conjugates, Eq.~(\ref{supermass}) contains
only $t$ mass terms of the forms ($i$) $M^{\vphantom{'}}_{t,i}
\bar{t}^{\vphantom{'}}_{L,i}t_{R,i}'$ for $i>1$, which transform as
$(\textbf{1},0)$ and are manifestly invariant under SU(2)$\times$U(1),
and ($ii$) $m_t \, \bar{t}_{L,i}t_{R,j}$, which originate via Yukawa
couplings that are also invariant under SU(2)$\times$U(1) when the
scalar fields are included, and which are made possible by the Higgs
VEV\@.  Therefore, any linear combination of these mass terms via
matrix diagonalization results in a gauge-invariant contribution to
$\mathcal{L}_{\rm mass}$.  In order to diagonalize the mass matrix,
one must solve a system more involved than the standard eigenvalue
problem in order to respect the metric $\rho$ of the LW Lagrangian.
To carry out the procedure, introduce symplectic transformations
$S_{L,R}$ for each supermultiplet $\Psi$ satisfying
\begin{equation} \label{quarkdiag}
  S_L^{\dagger} \, \rho \, S_L^{\vphantom{dagger}}=\rho,\;\;
  S_R^{\dagger} \, \rho \, S_R^{\vphantom{dagger}}=\rho,\;\;
  \mathcal{M} \rho=S_R^{\dagger} \,
  \mathcal{M}^{\vphantom{dagger}}_0\rho \,
  S_L^{\vphantom{dagger}} \, .
\end{equation}
Under these transformations, one identifies the linear combinations
$\Psi_{L,R}^{i,0} \equiv S_{L,R}^{ij} \Psi_{L,R}^j$ as the mass
eigenstates corresponding to the diagonal mass matrix
$\mathcal{M}^{\vphantom{dagger}}_0$, thereby providing each field
$\Psi_{L,R}^{i,0}$ with an unmixed propagator.  Appendix C of
Ref.~\cite{Figy:2011yu} provides an explicit procedure to rewrite
Eq.~(\ref{quarkdiag}) in terms of a standard Hermitian matrix
diagonalization.  The price of this diagonalization becomes apparent
in the kinetic terms.  Using the supermultiplet notation, one begins
by writing
\begin{equation} \label{Lkinetic}
  \mathcal{L}_{\rm kinetic} = \overline{T}_{\! L} \, i\cancel{D} \rho
  \, T_L + \overline{T}_{\! R} \, i\cancel{D} \rho \, T_R \, ,
\end{equation}
with an analogous expression for $T \to B$.  The calculation relevant
to this work requires one to include only the mass-eigenstate
$W^{\pm}_{i,0}$ bosons from the covariant derivative while ignoring
terms proportional to all the neutral gauge bosons, so that only the
portion of $i\cancel{D}$ in Eq.~(\ref{Lkinetic}) proportional to the
identity matrix in supermultiplet space is important here.
Furthermore, since the supermultiplets $T$ and $B$ contain fields with
different SM charges ({\it i.e.}, the primed {\it vs}.\ unprimed
fields), the gauge portion of $i\cancel{D}$ implicitly contains
projection operators (called $\Xi$ below) that eliminate
gauge-nonsinglet Lagrangian terms.  In order to implement the
transformation~(\ref{quarkdiag}), we adopt the notation
$T_{L,R}^i=\tau_{L,R}^{ij}T_{(L,R), \, 0}^{j}$, where
$\tau^{\vphantom{1}}_{L,R} \equiv S_{L, R}^{-1}$ of the $T$ sector.
In terms of the new basis, Eq.~(\ref{Lkinetic}) becomes:
\begin{equation} \label{Lkinetic0}
  \mathcal{L}_{\rm kinetic}
  =\overline{T}_{L,0}^{\vphantom{\dagger}} \tau_L^{\dagger} \,
  i\cancel{D} \rho \, \tau_L^{\vphantom{\dagger}}
  T_{L,0}^{\vphantom{\dagger}}
  +\overline{T}_{R,0}^{\vphantom{\dagger}} \tau_R^{\dagger} \,
  i\cancel{D} \rho \, \tau_R^{\vphantom{\dagger}}
  T_{R,0}^{\vphantom{\dagger}} \, .
\end{equation}
The analogous terms for the $B$ quarks are obtained introducing
$\beta_{L,R} \equiv S_{L,R}^{-1}$ for the $B$ sector and replacing
$\tau_L^{ij}T_{(L,R), \, 0}^j\rightarrow\beta_L^{ij}B_{(L,R), \, 0}^j$
in Eq.~(\ref{Lkinetic0}).  For the purposes of this calculation, the
most interesting element of the quark Lagrangian is also technically a
kinetic term, since it is derived from a covariant derivative:
\begin{eqnarray}
  \mathcal{L}_{\rm int} & = & \overline{T}_{L} \, i \cancel{D} B_L
  + {\rm h.c.} \nonumber \\ & \supset &
  \overline{T}^{\vphantom{\dagger}}_{L,0}\tau_L^{\dagger}
  \left(\frac{g_2}{\sqrt{2}}\hat{\theta}^{\vphantom{+}}_p V_{pq}^{+}
    \cancel{W}_{0,q}^{+}\right)\Xi_L^{\vphantom{\dagger}}
  \rho \, \beta_L^{\vphantom{\dagger}} B_{L,0}^{\vphantom{\dagger}}
  + {\rm h.c.} \label{Lint}
\end{eqnarray}
The CKM elements appear in this expression when one extends the
supermultiplets $T$ and $B$ to contain all 3 quark generations (and
would arise in part through the inequality of $\tau_L$ and $\beta_L$).
As described in Eq.~(\ref{thetadef}) and below, $\hat\theta$
incorporates the correlation of SM fields and LW partners from their
original HD superfields, and the matrix $V^{+}$ rotates the $W^+$
gauge bosons to their mass eigenstate basis.  The combined vector
$\hat{\theta}^{\vphantom{+}}_p V^+_{pq}$ comes from the form of the
fermion kinetic energy term $\bar{\psi}\gamma^{\mu}\hat{D}_{\mu}\psi$,
as in Ref.~\cite{Carone:2008iw}.  For $\psi$ transforming under the
fundamental representation of the group, the factor
$\hat{\theta}^{\vphantom{+}}_p V^+_{pq}$ attends the gauge boson
throughout the calculation; it becomes part of the gauge-fermion
vertex factor, and hence appears in the calculation of the decay rate.
For example, for LW $W$ masses $M_2=1$~TeV, $M_3=5$~TeV, one computes
$\hat{\theta}^{\vphantom{+}}_pV^+_{pq} \approx \{1, -1.021, 0.204 \}.$

$\Xi_L$ in Eq.~(\ref{Lint}) is defined as the matrix of 1's and 0's
that guarantees each nonvanishing term connecting components of $T_L$
and $B_L$ is a gauge singlet [Likewise, for the gauge boson $B$, the
1's would be replaced with U(1) charges].  An analogous
$\mathcal{L}_{\rm int}$ can be formed from the right-handed mass
eigenstates.  One thus defines the generalized quark mixing matrices:
\begin{equation}
  \Lambda^{\vphantom{\dagger}}_{L,R} \equiv \tau_{L,R}^{\dagger} \,
  \Xi^{\vphantom{\dagger}}_{L,R} \, \rho \,
  \beta^{\vphantom{\dagger}}_{L,R} \, .
\end{equation}
With this convenient abbreviation, one can employ the usual chirality
projection operators $P_{L,R} \equiv \frac 1 2 (1 \mp
\gamma_5)$ to write $\Psi_0 \equiv \Psi_{L,0} P_L + \Psi_{R,0} P_R$
and cast the full interaction term as
\begin{equation}
  \mathcal{L}_{\rm int} =  \frac{g_2}{\sqrt{2}}
  \overline{T}^{\vphantom{\dagger}}_0
  \hat{\theta}_p^{\vphantom{+}} V_{pq}^{+} \cancel{W}_{0,q}^{+}
  (\Lambda^{\vphantom{\dagger}}_L P^{\vphantom{\dagger}}_L
  +\Lambda^{\vphantom{\dagger}}_R P^{\vphantom{\dagger}}_R)
  B^{\vphantom{\dagger}}_0 + {\rm h.c.} \label{LintTB}
\end{equation}

\subsection{$W$ Boson Width Calculation}

In the special case of $W^+\rightarrow t\bar{b}$ decay, the associated
Feynman vertex rule from Eq.~(\ref{LintTB}) reads
\begin{equation}
  i\frac{g_2}{\sqrt{2}}\gamma^{\mu}\hat{\theta}_p^{\vphantom{+}}
  V_{pq}^{+} (\Lambda_L^{ij} P^{\vphantom{+}}_L
  + \Lambda^{ij}_R P^{\vphantom{+}}_R) \, ,
\end{equation}
which leads to the invariant matrix element
\begin{equation}
i\mathcal{M}=i\epsilon_{\mu}\frac{g_2}{\sqrt{2}}\bar{t}_0^{\, i}
\gamma^{\mu}\hat{\theta}^{\vphantom{+}}_p V_{pq}^{+}
(\Lambda_L^{ij} P^{\vphantom{j}}_L
+\Lambda_R^{ij}P^{\vphantom{j}}_R)b_0^{\, j} \, ,
\end{equation}
and then to the squared, spin-averaged matrix element
\begin{eqnarray}
  \overline{|\mathcal{M}|^2} & = & \frac{g_2^2}{3}
  |\hat{\theta}_pV_{pq}^{+}|^2 \! \left\{ \left[ M_{W,q}^2
  -\frac{1}{2} (m_{t,i}^2+m_{b,j}^2) -\frac{1}{2M_{W,q}^2}
  (m_{t,i}^2-m_{b,j}^2)^2 \right]
  (\Lambda_L^{ij}\Lambda_L^{\dagger ji}+\Lambda_R^{ij}
  \Lambda_R^{\dagger ji}) \right. \nonumber \\ & + &
  \left. 3 m_{t,i} m_{b,j} (\Lambda_L^{ij}\Lambda_R^{\dagger ji}
  +\Lambda_R^{ij}\Lambda_L^{\dagger ji}) \right\} \, .
\label{tbelement}
\end{eqnarray}
No Einstein summation is assumed on the indices $q$, $i$, $j$, so that
Eq.~(\ref{tbelement}) specifies the squared amplitude for the $q^{\rm
  th}$ weak gauge boson, the $i^{\rm th}$ top quark state, and the
$j^{\rm th}$ bottom quark state (all mass eigenstates).  In the SM
case, $|\hat{\theta}^{\vphantom{+}}_pV_{pq}^{+}|^2 = 1$, $\Lambda_L =
V_{tb}$, and $\Lambda_R = 0$.

One then integrates over phase space to obtain the decay width
$\Gamma$.  Using the well-known formulas
\begin{equation}
  \Gamma=\frac{1}{2M_{W,q}}\int d\Pi_2|\mathcal{M}|^2 \, ,
\end{equation}
where
\begin{equation}
  \int d\Pi_2=\int d \cos\theta \, \frac{1}{16\pi}\frac{2|\textbf{p}|}
  {E_{cm}}=\frac{1}{4\pi M_{W,q}}\sqrt{
  \frac{(M_{W,q}^2+m_{t,i}^2-m_{b,j}^2)^2}{4M_{W,q}^2}-m_{t,i}^2} \, ,
\end{equation}
one finds the total contribution to the width of the $q^{\rm th}$ weak
gauge boson to be
\begin{equation}
\Gamma = \frac{\overline{|\mathcal{M}|^2}}{8\pi M_{W,q}^2}\sqrt{
\frac{(M_{W,q}^2+m_{t,i}^2-m_{b,j}^2)^2}{4M_{W,q}^2}-m_{t,i}^2} \, ,
\end{equation}
assuming of course that $M_{W,q} > m_{t,i} + m_{b,j}$.

Working under the assumption that the final-state SM quarks are
essentially massless with respect to the initial-state LW gauge boson,
the decay rate contribution for each $W^+_q\rightarrow f_i\bar{f}_j$
event is
\begin{equation} \label{decaycont}
  \delta \Gamma =g_2^2\frac{|\hat{\theta}_p V^{+}_{pq}|^2}{48\pi}
  (\Lambda_L^{ij} \Lambda_L^{\dagger ji}
  +\Lambda^{ij}_R\Lambda_R^{\dagger ji})M^{\vphantom{\dagger}}_{W,q}
  \, ,
\end{equation}
For the case $M_3\gg M_2,$ one anticipates from Eq.~(\ref{thetadef})
that $|\hat{\theta}^{\vphantom{+}}_pV^+_{p2}|
\gg|\hat{\theta}^{\vphantom{+}}_pV^+_{p3}|$, which suppresses the
decay rate contribution for $W_3^+\rightarrow f_1\bar{f}_2$ compared
to that for $W^{+}_2$.  This effect is mitigated by the possible
presence of massive final-state particles kinematically forbidden in
$W^{+}_2$ decays but allowed in $W^{+}_3$ decays.

\begin{acknowledgments}
This work was supported in part by the National Science Foundation
under Grant Nos.\ PHY-0757394 and PHY-1068286.
\end{acknowledgments}


\begin{thebibliography}{99}

\bibitem{Grinstein:2007mp} 
  B.~Grinstein, D.~O'Connell, and M.B.~Wise,
  Phys.\ Rev.\ D {\bf 77}, 025012 (2008)
  [arXiv:0704.1845 [hep-ph]].

\bibitem{Lee:1970iw} 
  T.D.~Lee and G.C.~Wick,
  Phys.\ Rev.\ D {\bf 2}, 1033 (1970).

\bibitem{Rizzo:2007ae}
  T.G.~Rizzo,
  JHEP {\bf 0706}, 070 (2007)
  [arXiv:0704.3458 [hep-ph]].

\bibitem{Alvarez:2008za} 
  E.~Alvarez, L.~Da~Rold, C.~Schat, and A.~Szynkman,
  JHEP {\bf 0804}, 026 (2008) [arXiv:0802.1061 [hep-ph]].

\bibitem{Carone:2008bs} 
  C.D.~Carone and R.F.~Lebed,
  Phys.\ Lett.\ B {\bf 668}, 221 (2008)
  [arXiv:0806.4555 [hep-ph]].

\bibitem{Underwood:2008cr}
  T.E.J.~Underwood and R.~Zwicky,
  Phys.\ Rev.\ D {\bf 79}, 035016 (2009)
  [arXiv:0805.3296 [hep-ph]].
  
\bibitem{Chivukula:2010nw} 
  R.S.~Chivukula, A.~Farzinnia, R.~Foadi, and E.H.~Simmons,
  Phys.\ Rev.\ D {\bf 81}, 095015 (2010)
  [arXiv:1002.0343 [hep-ph]].

\bibitem{Krauss:2007bz} 
  F.~Krauss, T.E.J.~Underwood, and R.~Zwicky,
  Phys.\ Rev.\ D {\bf 77}, 015012 (2008) [Erratum-ibid.\ D {\bf 83},
  019902 (2011)] [arXiv:0709.4054 [hep-ph]].

\bibitem{Carone:2009nu}
  C.D.~Carone and R.~Primulando,
  Phys.\ Rev.\ D {\bf 80}, 055020 (2009) [arXiv:0908.0342 [hep-ph]].

\bibitem{Alvarez:2011ah} 
  E.~Alvarez, E.C.~Leskow, and J.~Zurita,
  Phys.\ Rev.\ D {\bf 83}, 115024 (2011) [arXiv:1104.3496 [hep-ph]].

\bibitem{Figy:2011yu} 
  T.~Figy and R.~Zwicky,
  JHEP {\bf 1110}, 145 (2011) [arXiv:1108.3765 [hep-ph]].

\bibitem{Carone:2008iw} 
  C.D.~Carone and R.F.~Lebed,
  JHEP {\bf 0901}, 043 (2009)
  [arXiv:0811.4150 [hep-ph]].

\bibitem{Pais:1950za} 
  A.~Pais and G.E.~Uhlenbeck,
  Phys.\ Rev.\ {\bf 79}, 145 (1950).

\bibitem{LTprep}
  R.F.~Lebed and R.H.~TerBeek, in preparation.

\bibitem{Melnikov:2006kv} 
  K.~Melnikov and F.~Petriello,
  Phys.\ Rev.\ D {\bf 74}, 114017 (2006) [hep-ph/0609070].

\bibitem{Martin:2009iq} 
  A.D.~Martin, W.J.~Stirling, R.S.~Thorne, and G.~Watt,
  Eur.\ Phys.\ J.\ C {\bf 63}, 189 (2009)
  [arXiv:0901.0002 [hep-ph]].
	 
\bibitem{Davoudiasl:2007cy} 
  H.~Davoudiasl and T.G.~Rizzo,
  Phys.\ Rev.\ D {\bf 76}, 055009 (2007)  [hep-ph/0702078 [HEP-PH]].

\bibitem{Barbieri:2004qk} 
  R.~Barbieri, A.~Pomarol, R.~Rattazzi, and A.~Strumia,
  Nucl.\ Phys.\ B {\bf 703}, 127 (2004)  [hep-ph/0405040].


\bibitem{ArkaniHamed:1999za} 
  N.~Arkani-Hamed, Y.~Grossman, and M.~Schmaltz,
  Phys.\ Rev.\ D {\bf 61}, 115004 (2000) [hep-ph/9909411].

\bibitem{Rizzo:1999br} 
  T.G.~Rizzo and J.D.~Wells,
  Phys.\ Rev.\ D {\bf 61}, 016007 (2000)
  [hep-ph/9906234].

\end{thebibliography}
\end{document}